\documentclass[12pt]{article}
\usepackage{latexsym,graphicx,amssymb,color,amsmath}
\usepackage{lscape}
\newcommand{\be}{\begin{equation}}
\newcommand{\ee}{\end{equation}}
\newcommand{\ba}{\begin{eqnarray}}
\newcommand{\ea}{\end{eqnarray}}

\begin{document}
\vspace*{-2cm}
\begin{flushright}
DCPT-07/67\\
\end{flushright}

\vspace{0.3cm}

\begin{center}
{\Large {\bf %RNA and 
DNA Duplex Cage Structures with Icosahedral Symmetry}}\\ 
\vspace{1cm} {\large  \bf N. E.\
Grayson\,\footnote{\noindent E-mail: {\tt neg100@york.ac.uk}}, A.\
Taormina\,\footnote{\noindent E-mail: {\tt anne.taormina@durham.ac.uk}} and 
R.\ Twarock\,\footnote{\noindent E-mail: 
{\tt rt507@york.ac.uk}}}\\
\vspace{0.3cm} {${}^{1,3}$}\em Department of Mathematics and Department of Biology\\ University of York\\
York YO10 5DD, U.K.\\
 \vspace{0.3cm} {${}^{2}$\em \it Department of Mathematical
Sciences\\ University of Durham\\ Durham DH1 3LE, U.K.}\\ 
\end{center}

\begin{abstract}
A construction method for duplex cage structures with icosahedral symmetry made out of single-stranded  
DNA molecules is presented and applied to an icosidodecahedral cage. 
It is shown via a  mixture of analytic and computer techniques that there exist realisations of this graph in terms of two circular DNA molecules. These blueprints for the organisation of a cage structure with a noncrystallographic symmetry may assist in the design of containers made from DNA for applications in nanotechnology. 
\end{abstract}

\section{Introduction}
RNA cages are known to occur in certain families of viruses. For example, a proportion of the viral RNA of Pariacoto virus is packaged within the viral particles in the form of a dodecahedral RNA cage \cite{Tang}, and bacteriophage MS2 is known to package part of its genomic material in the form of a 32-faced polyhedron reminiscent of the buckyball 
\cite{Ranson,Worm}. Remarkably, recent advances in biotechnology provide the necessary tools to engineer
cage structures from nucleic acids, and open novel avenues for applications in nanotechnology. 

{DNA} cages with crystallographic symmetry have already been realised experimentally in the shape of a cube \cite{cube}, a tetrahedron \cite{tetrahedron}, an octahedron \cite{octahedron} or a truncated octahedron \cite{truncatedoctahedron}, and one natural idea is to use such cages for cargo delivery or storage \cite{drug}. A systematic, theoretical analysis of %RNA and 
DNA cage structures is still lacking, and our motivation here is the hope our mathematical considerations on the organisation of  %RNA or 
DNA  in cages with icosahedral symmetry will aid the design of artificial cages inspired by nature. 

Models for dodecahedral cages have been derived in \cite{Bruinsma, JT}. Since a dodecahedron has trivalent vertices  and a small number of faces, the combinatorics involved in solving the %RNA 
{DNA} organisation problem can be done without computer help. This is no longer the case for a four-coordinated polyhedron with thirty-two faces such as the icosidodecahedron. However, icosidodecahedral cages are of particular interest because their volume per surface ratio is the largest in comparison with dodecahedral and icosahedral cages, making them the more appropriate option among these noncrystallographic cages for applications in which the storage or transportation of a larger cargo is required.

We start by introducing our theoretical construction method in general terms for all polyhedra
with icosahedral symmetry in Section \ref{sec1}, and then concentrate on icosidodecahedral cages in Section  \ref{sec2}  via an approach that combines symmetry arguments and computer analyses.

\section{The general set-up: Orientable embeddings and DNA cage structures}
\label{sec1}

We consider the organisation of circular single-stranded DNA (ssDNA) molecules on cages
with icosahedral symmetry as in \cite{Worm} \footnote{dodecahedral, icosahedral, 32-faceted polyhedral.}, such that every edge is met by a strand precisely twice in opposite directions. This rule ensures that two different portions of the strand meeting along an edge may hybridize into a duplex structure with the two strands oriented in opposite 3' to 5' directions along that edge.

 From a mathematical point of view, we consider the cage as being a graph $G$ whose nodes are the vertices of the cage, and the connectors are its edges. We then search for {\em orientable thickened graphs}, which are  compact orientable 2-dimensional surfaces constructed out of strips and thickened $n$-junctions glued together, such that the original graph $G$ is topologically embedded into such thickened graphs as a deformation retract  \cite{Jonoska,Greenberg}. The boundary curves of these thickened graphs form part of circular ssDNA molecules. The aim of this analysis is to realize the graph in terms of a minimal number of circular strands which visit each edge of the cage  twice in opposite directions so that the strand segments may hybridize into double helical structures along the edges. 

\begin{figure}[ht]
\begin{center}
\raisebox{-1.3cm}{\includegraphics[width=8cm,keepaspectratio]{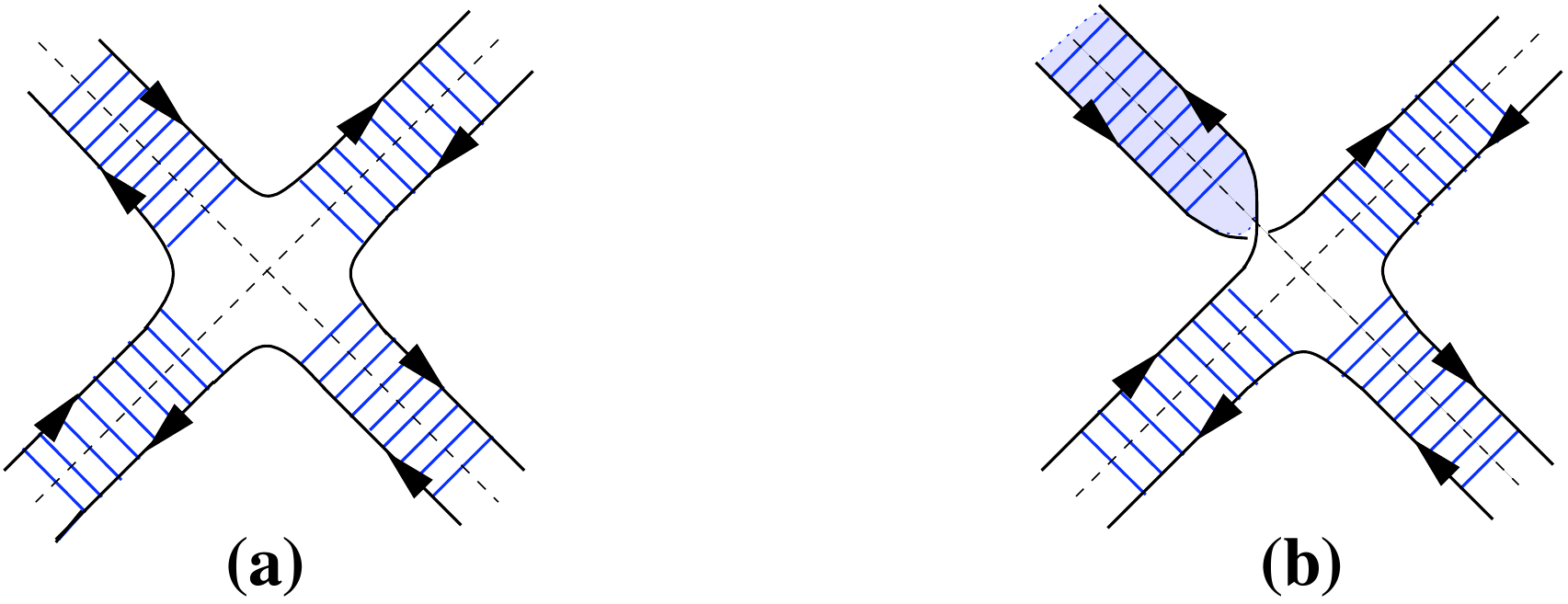}}
\end{center}
\caption{{\em Thickened 4- junction of type A (a) and of type B (b). The line segments represent base pairs.}}
\label{OptimalVertex}
\end{figure}

We present an optimization procedure, which takes the following factors into account.
\begin{enumerate}
\item {\em Initial data}: Assuming the cages are made of polygons with all sides of equal length $\lambda$, one can imagine to manufacture ssDNA  cages of different sizes and therefore,  the number $\nu(\lambda)$ of half-turns in the duplex structure along the edges depends on $\lambda$ \footnote{For viruses observed in vivo,
the RNA cages have specific sizes and therefore a well-defined $\nu(\lambda).$}. Configurations where $\nu(\lambda)$ is odd are  modelled as cross-overs in the planar projective views of the polyhedra that we are using for our analysis. We remark that there are about 10.5 base pairs (bp) per helical turn. 

\item {\em Thickened $n$-junctions}: Mechanical stress may be imposed on the overall configuration if junctions with an extra number of twists  (helical turns) on the legs of the thickened $n$-junction are introduced. For example, the thickened 4-junction shown in Fig.~\ref{OptimalVertex}(a) imposes no stress on the configuration (we name it `type A'), whilst the thickened 4-junction appearing in Fig.~\ref{OptimalVertex}(b) accommodates one single twist (we name it `type B') and imposes stress on the overall configuration unless extra nucleotides are introduced (in the non-basepaired inner part of the junction) that compensate for it.
\end{enumerate}
%----------------------------------------- Figure --------------------
%--------------------------------------------------------------------
The number $n$ of legs in the junctions depends on the type of cage considered. In Section 3, we consider cages having 4-junctions\footnote{Four-junctions can be manufactured, see for example \cite{Holliday}.}, but 
for dodecahedral cages for instance, 3-junctions are needed. In
\cite{Bruinsma}, different types of 3-junctions are studied. Apart from those similar to the 4-junctions in Fig.~\ref{OptimalVertex}, they  consider 3-junctions that involve the occurrence of hairpins. We exclude such junctions from our analysis here because they require an extra discussion of how to keep the hairpin in place. In the context of viruses, for example, the protein container acts as a scaffold for the RNA cages. In the context of nanocontainers where such a scaffold is lacking, the hairpin has to be designed such that it interacts with the remaining part of the junction, or, is attached via a sticker strand \cite{Rothemund}.  

The first step in the optimization procedure is to identify {\em start configurations}, i.e. orientable thickened graphs with a maximum of Type A thickened junctions. Such graphs are usually made of several distinct circular strands 
that we call {\it loops}. Our ultimate goal is to construct the graph from a minimal number of such loops, and therefore further junctions need to be replaced in the start configuration to merge loops in a next step.

In order to determine the start configurations, we start by assuming that every vertex on the polyhedron is represented by a type A thickened $n$-junction. However, in the presence of cross-overs which take into account the odd  number of half-turns along the edges, this distribution of type A junctions 
does not necessarily provide  an orientable thickened graph. In particular, this is the case if faces of a cage with an odd number of cross-overs occur. 

\begin{figure}[ht]
\begin{center}
(a) \raisebox{1cm}{\includegraphics[width=4cm,keepaspectratio]{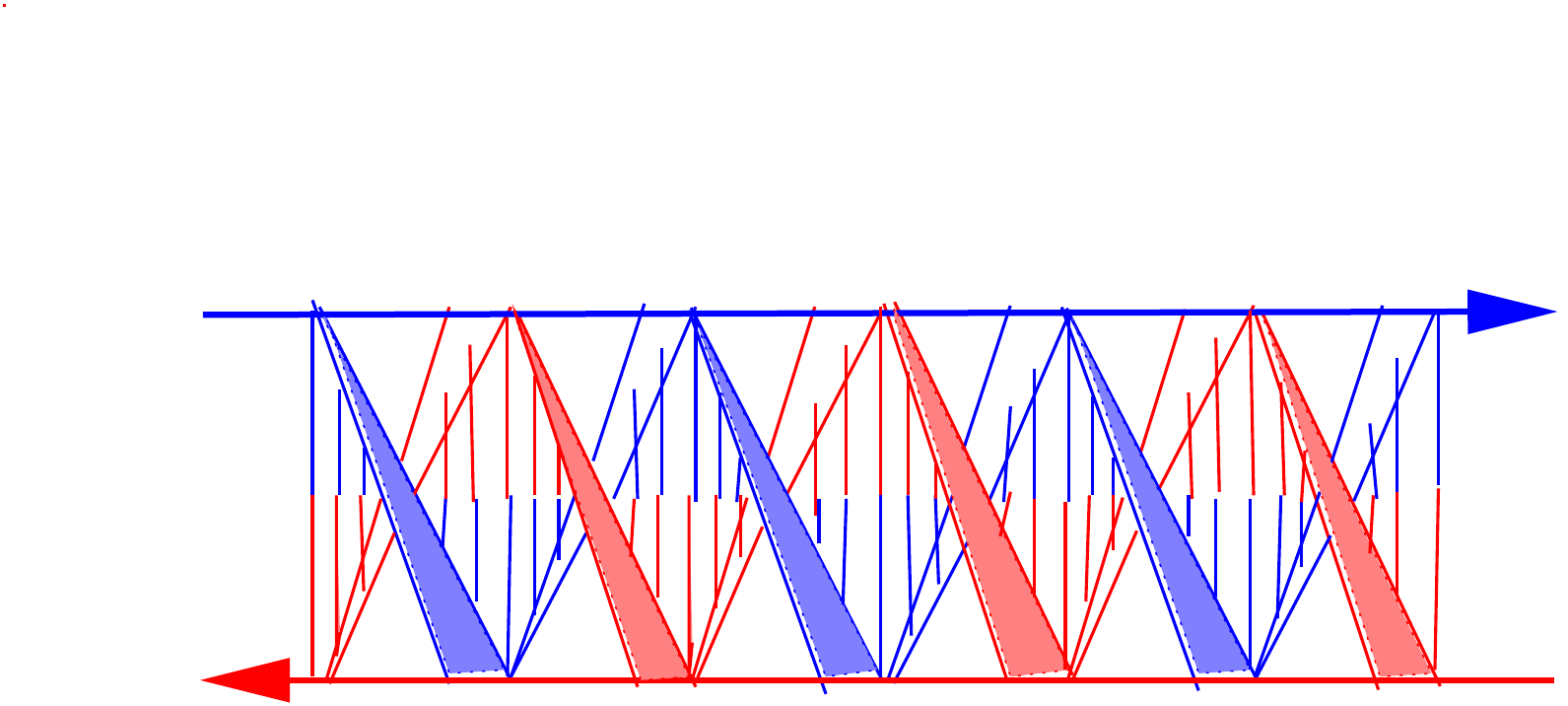}}
\raisebox{1.5cm}{$\longrightarrow$}\quad\includegraphics[width=4cm,keepaspectratio]{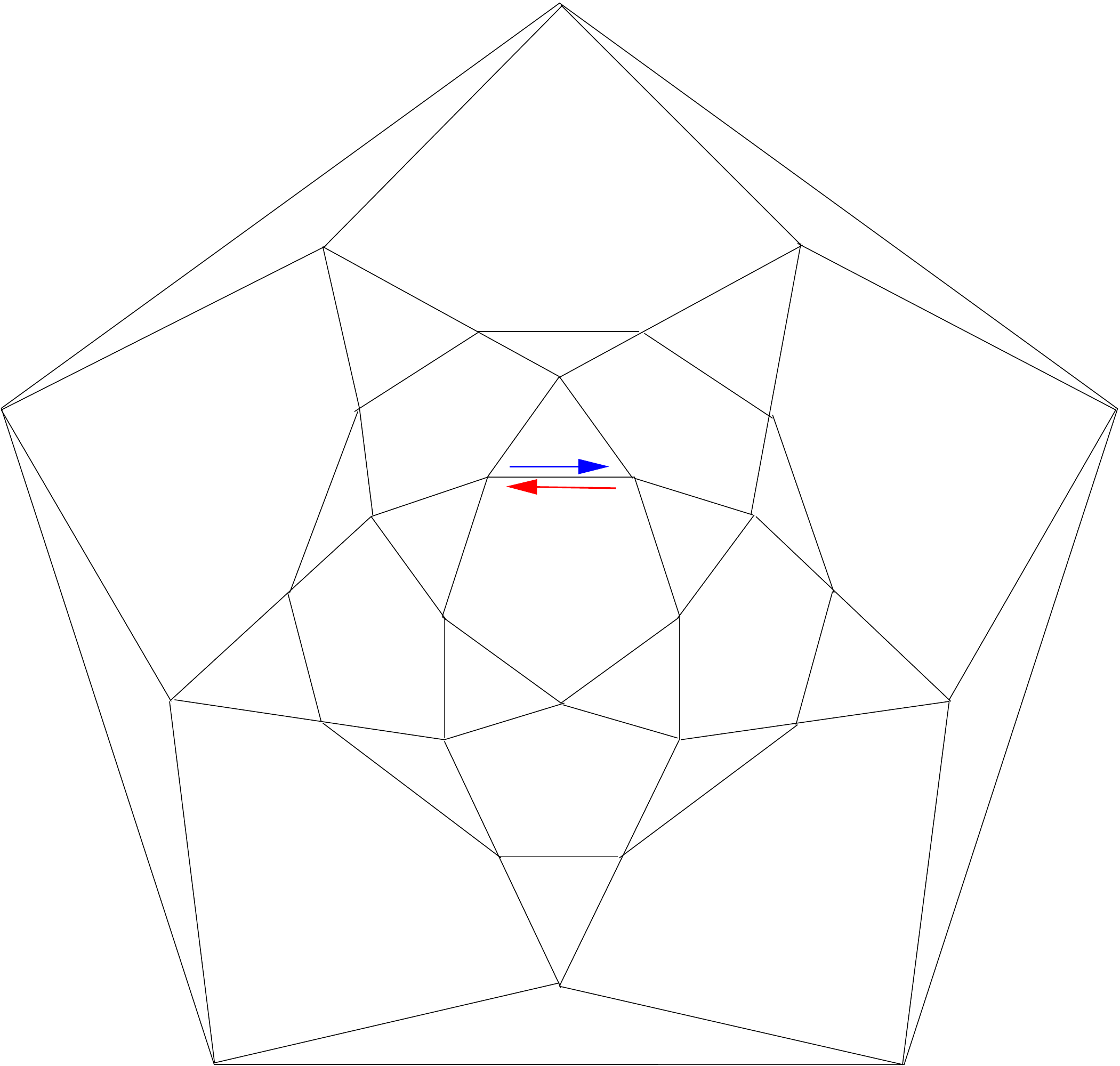}\\
(b) \raisebox{1cm}{\includegraphics[width=4cm,keepaspectratio]{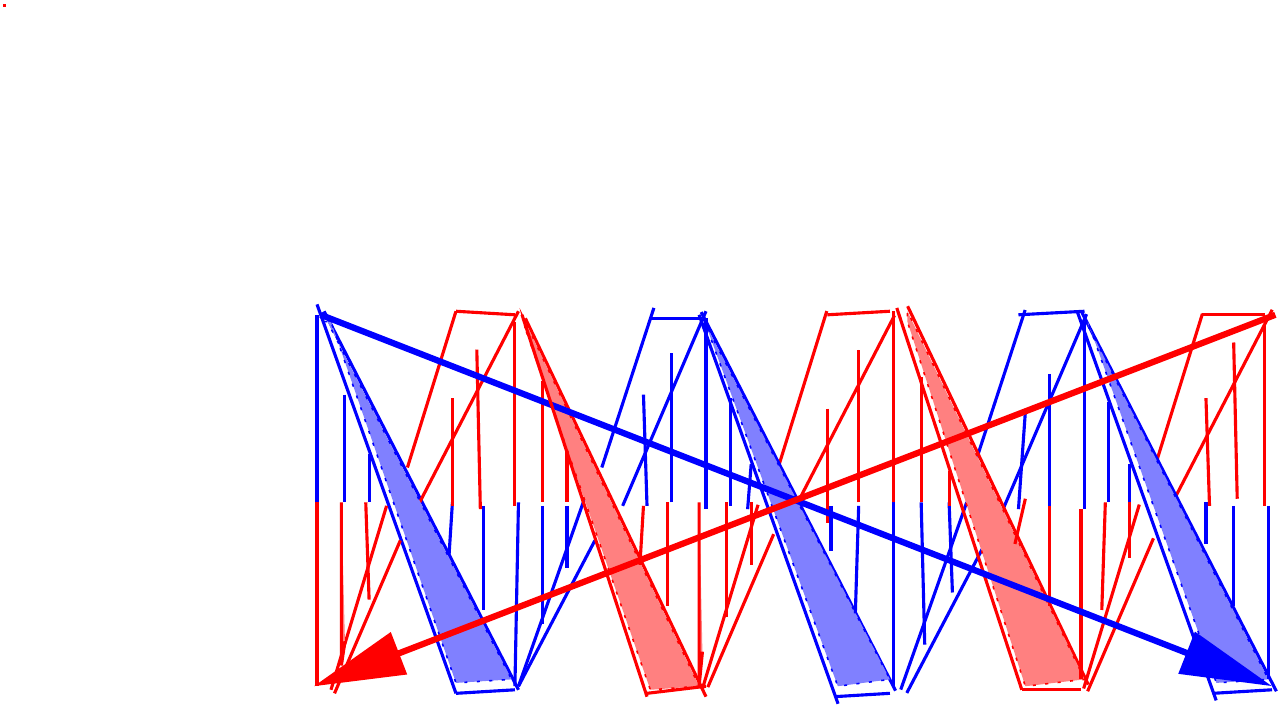}}
\raisebox{1.5cm}{$\longrightarrow$}\quad\includegraphics[width=4cm,keepaspectratio]{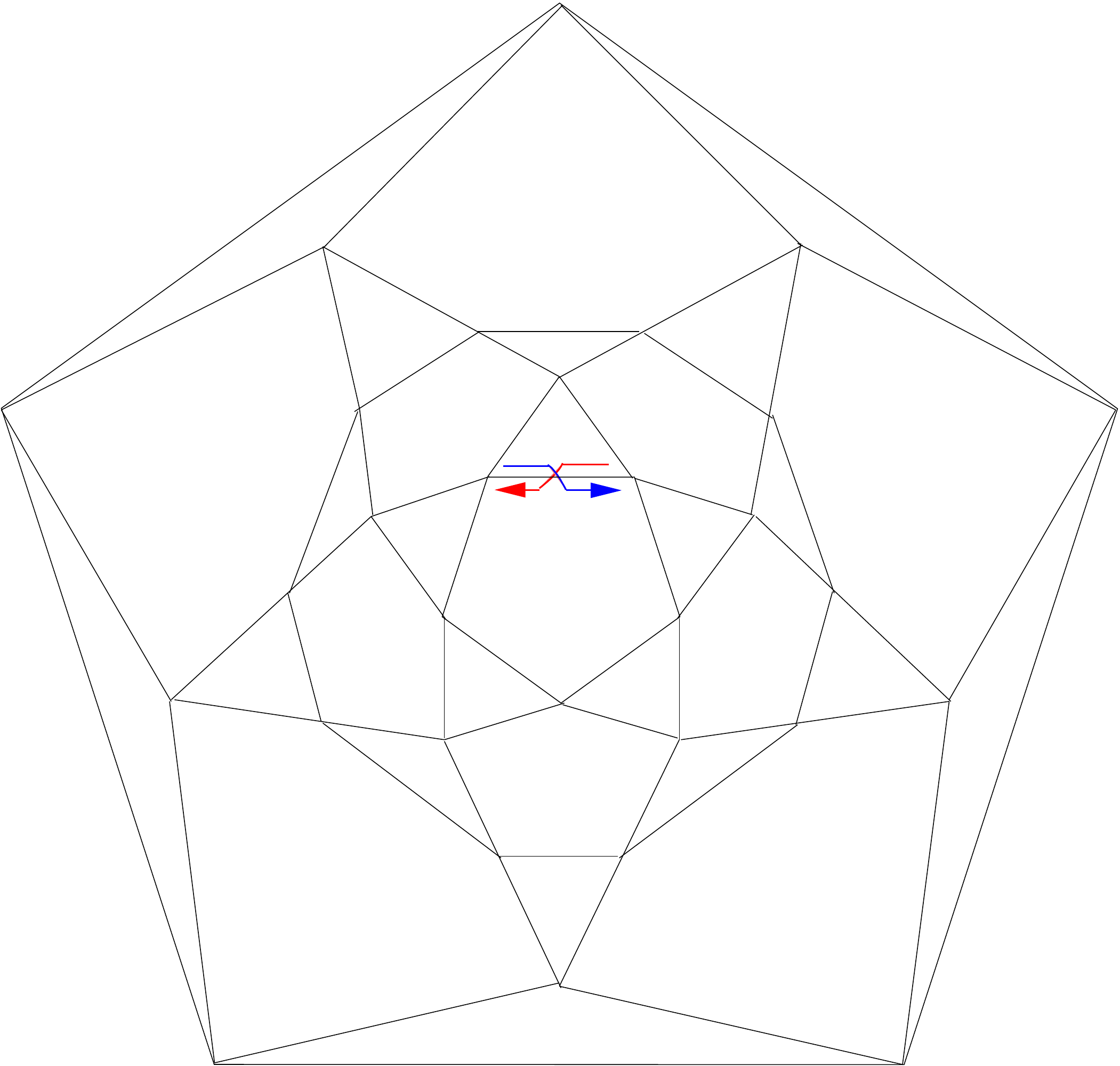}
\end{center}
\caption{{\em (a) The double helix of %RNA or 
DNA is represented by lines (blue and red) that trace the backbone of the helices. (b) Depending on their lengths, additional half-turns may appear, that are represented by cross-overs on the planar representation of the graph.} }
\label{options}
\end{figure}

In this paper, we restrict ourselves to cages which have all edges of the same length, so that the number of helical turns is the same on all edges to start with. Therefore {\em either} the strips along these edges are not twisted, i.e. all edges are cross-over free, {\em or} the strips are twisted, and all edges have cross-overs, see Fig.~\ref{options}. 
%----------------------------------------- Figure --------------------
%--------------------------------------------------------------------

\begin{figure}[ht]
\begin{center}
\raisebox{-1.3cm}{\includegraphics[width=6cm,keepaspectratio]{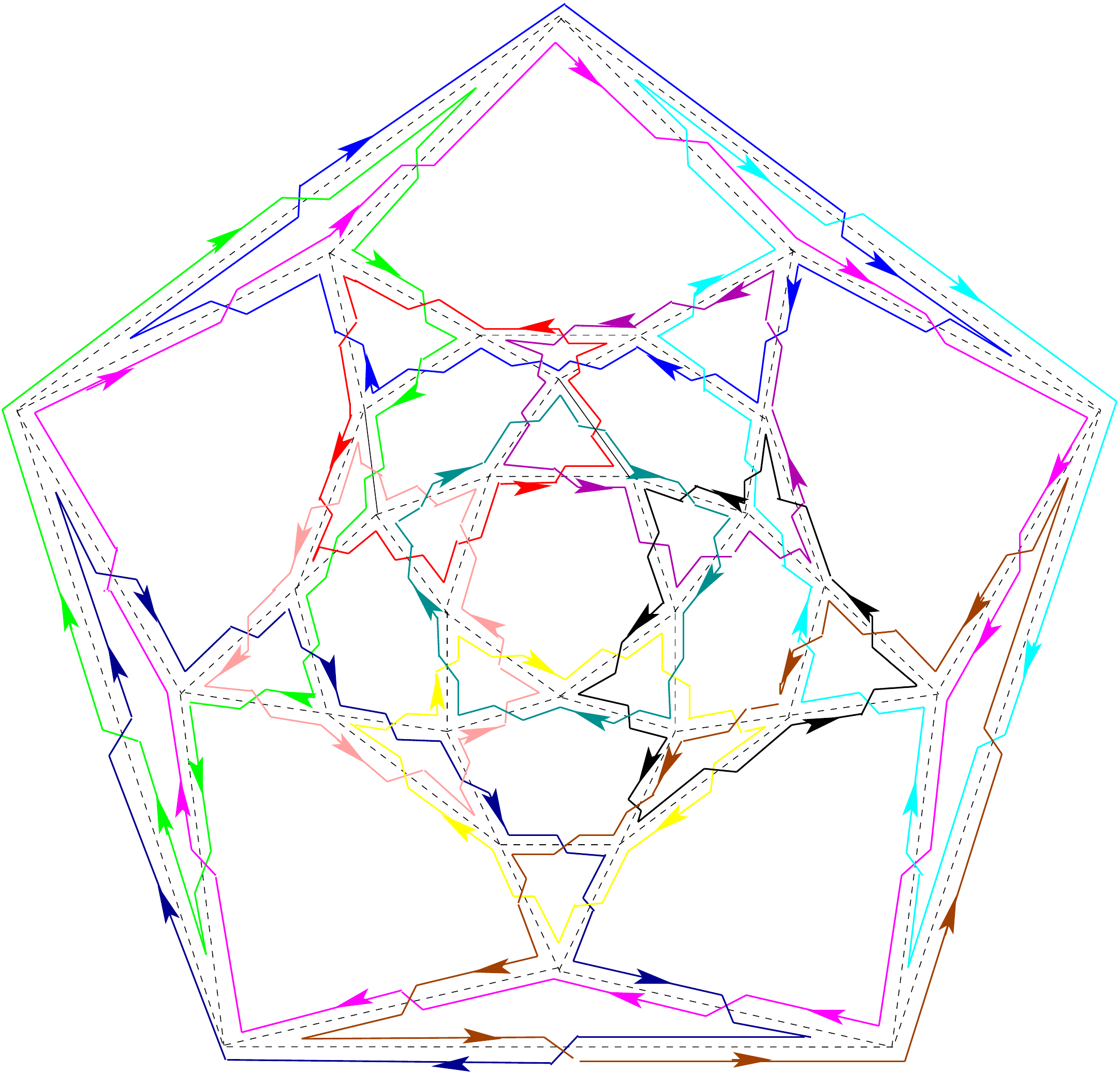}}
\end{center}
\caption{{\em Initial data configuration for the icosidodecahedron cage when $\nu(\lambda)$ is odd. }}
\label{initialdata}
\end{figure}

\begin{figure}[ht]
\begin{center}
\raisebox{-1.3cm}{\includegraphics[width=8cm,keepaspectratio]{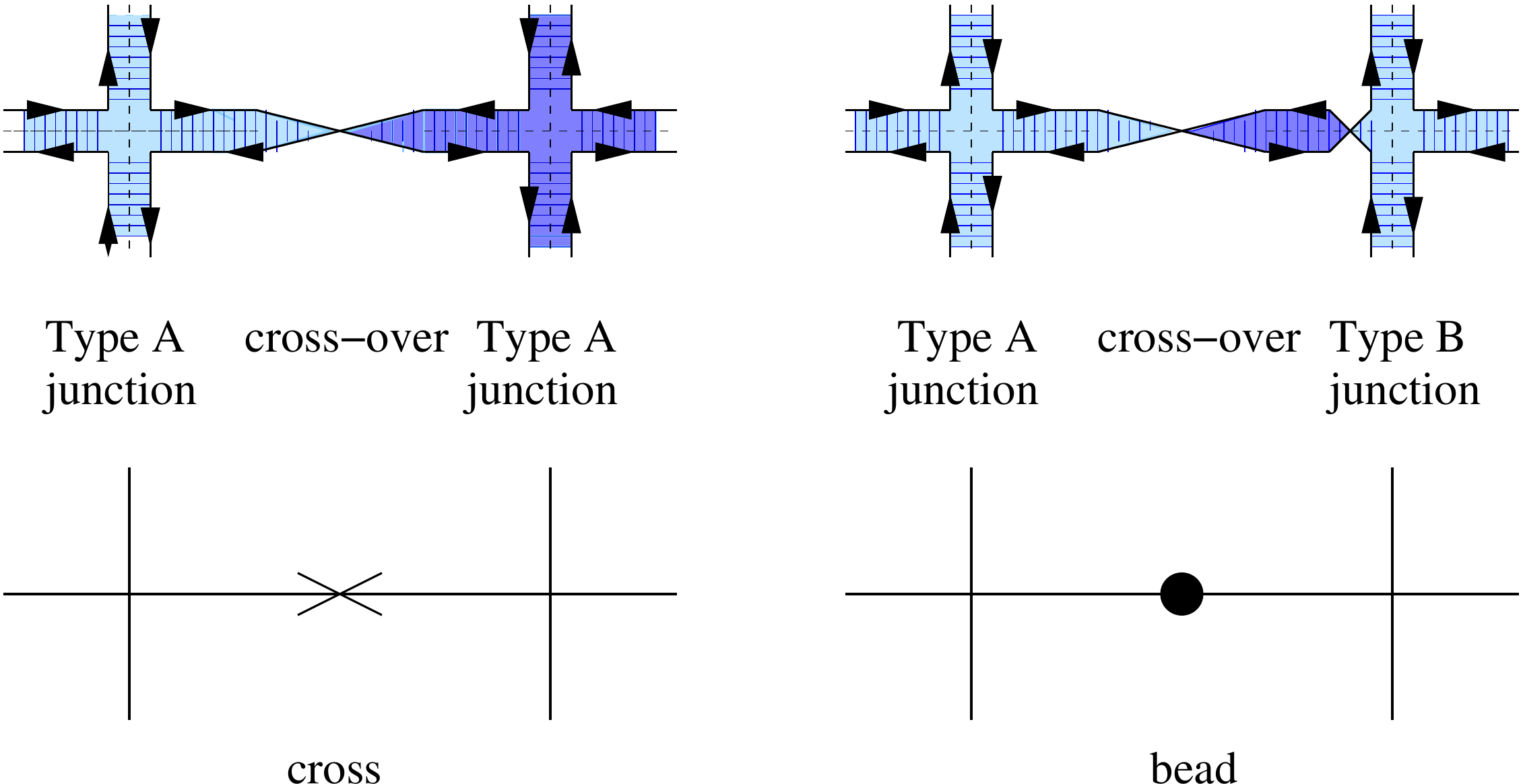}}
\end{center}
\caption{{\em Emergence of a bead in a modified initial data configuration. }}
\label{bead}
\end{figure}

In the first case, the start configuration is straightforward: All junctions are of type A, and there are as many loops as faces of the polyhedron considered. In the second case where all edges exhibit cross-overs as a consequence of their length, we call {\em initial  data configuration} the 2d surface (orientable or not) obtained by gluing the twisted strips representing the cross-overs to type A thickened $n$-junctions.  Fig.~\ref{initialdata} shows the initial data configuration of the icosidodecahedron. This configuration has twelve loops, but they do not all run in opposite directions so that the initial data configuration does not provide a suitable template for a DNA cage.
%----------------------------------------- Figure --------------------
%--------------------------------------------------------------------
In order to decide how many junctions of type B must be introduced in the initial data configuration to obtain a start configuration, we use the notion of {\em bead} introduced in \cite{JT}, and follow the {\em bead rule}. A bead appears on an edge of the polyhedron whenever a twisted strip (cross-over) is glued to the twisted leg of a type B thickened junction, as examplified in Fig.~\ref{bead} for a 4-coordinated polyhedron.
\vskip .5cm
%----------------------------------------- Figure --------------------
%--------------------------------------------------------------------

The bead rule requires the placement of beads on the edges of the polyhedron such that {\it all} of the following  three conditions are satisfied: 
\begin{itemize}
\item Each edge accommodates either a cross-over or a bead.
\item Every face of the polyhedron in the start configuration must have an {\it even} number of cross-overs.
\item The number of beads is minimal.
\end{itemize}

After the bead rule has been applied, one must identify configurations which are equivalent under icosahedral symmetry. The symmetry-inequivalent bead configurations correspond to the possible start configurations, where some junctions are of type A, and others of type B, the latter having been introduced to provide orientability of the 2d surface. 

\begin{figure}[ht]
\begin{center}
\includegraphics[width=10cm,keepaspectratio]{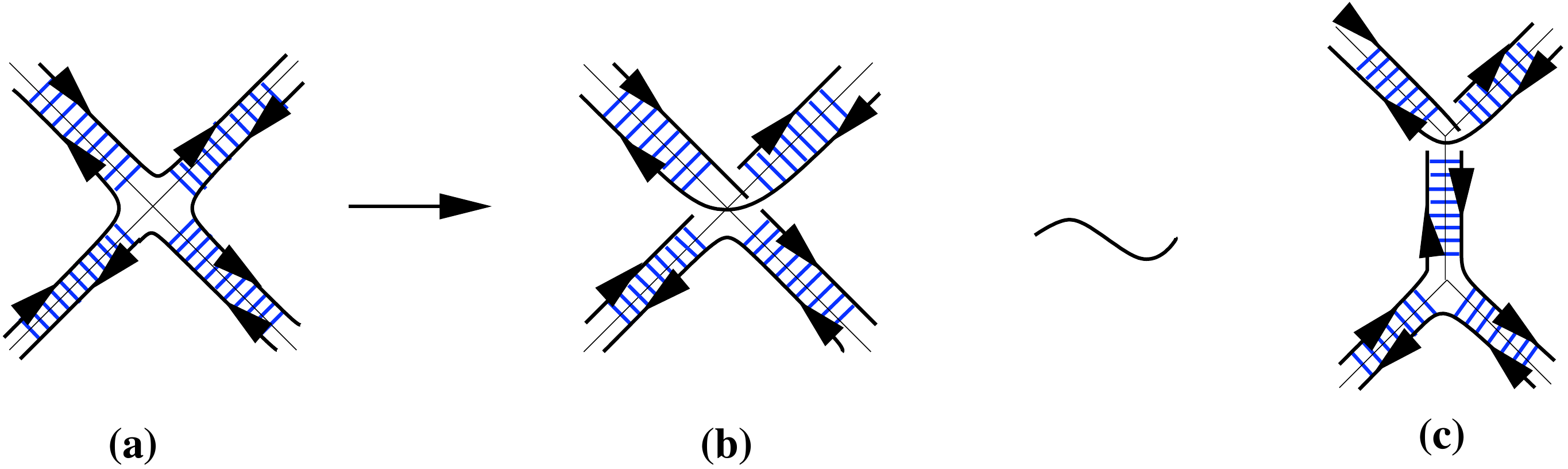}
%(a)\,\,\includegraphics[width=6cm,keepaspectratio]{3to1.pdf}\qquad
%(b)\,\,\includegraphics[width=6cm,keepaspectratio]{3to1bis.pdf}
\end{center}
\caption{{\em %(a) Replacement of a 4-loop configuration by a single loop
%(Type I replacement), 
(a) Type A 4-junction, (b) Replacement 4-junction which allows to merge 4 loops into 2, or 3 loops into a single one, (c)  The replacement 4-junction is equivalent to two 3-junctions.}}
\label{replace}
\end{figure}

In order to realize cages via a minimal number of circular DNA molecules, further junctions have to be replaced  in the start configuration in order to merge several individual loops into a single loop. We will use the replacement shown in { Fig.~\ref{replace}(b)}, which is the 4-junction analogue to the 3-junction replacement used in \cite{JT}. 
%---------------------------------------- figure --------------------
%--------------------------------------------------------------------
Such a replacement results in the reduction of the overall number of loops (for example in the start configuration) by two. To see this, observe that after replacement the 4-junction is topologically equivalent to two 3-junctions as shown on {Fig.~\ref{replace}(c)}. The upper 3-junction corresponds to Fig.~3 in \cite{JT} and, according to this reference, changes the number of loops by two, whilst the lower 3-junction has no effect on the number of loops. The 4-junction replacement thus reduces the number of loops by two. 
%Since the  reduction in the number of loops caused by this 4-junction replacement is a combination of the effects due to these two 3-junctions, it is hence also two.
 Moreover, note that the upper 3-junction differs from the trivial junction by a few nucleotides only, and hence this 4-junction does not impose stress on the overall cage structure if, { as discussed in \cite{JT}}, %according to \cite{JT}, if 
 a few extra nucleotides are introduced at the junctions. 

\section{The example of the icosidodecahedral cage} 
\label{sec2}

We consider the 32-faced polyhedron with thirty quatrovalent vertices as an example. Since all edges are of the same length, there are two different scenarios to consider, depending on the sizes of the DNA molecules, see Fig.~\ref{options}. {These scenarios are discussed below.}

\subsection{%None of the edges has an additional cross-over.
Cross-over free cages}

\begin{figure}[ht]
\begin{center}
\includegraphics[width=7cm,keepaspectratio]{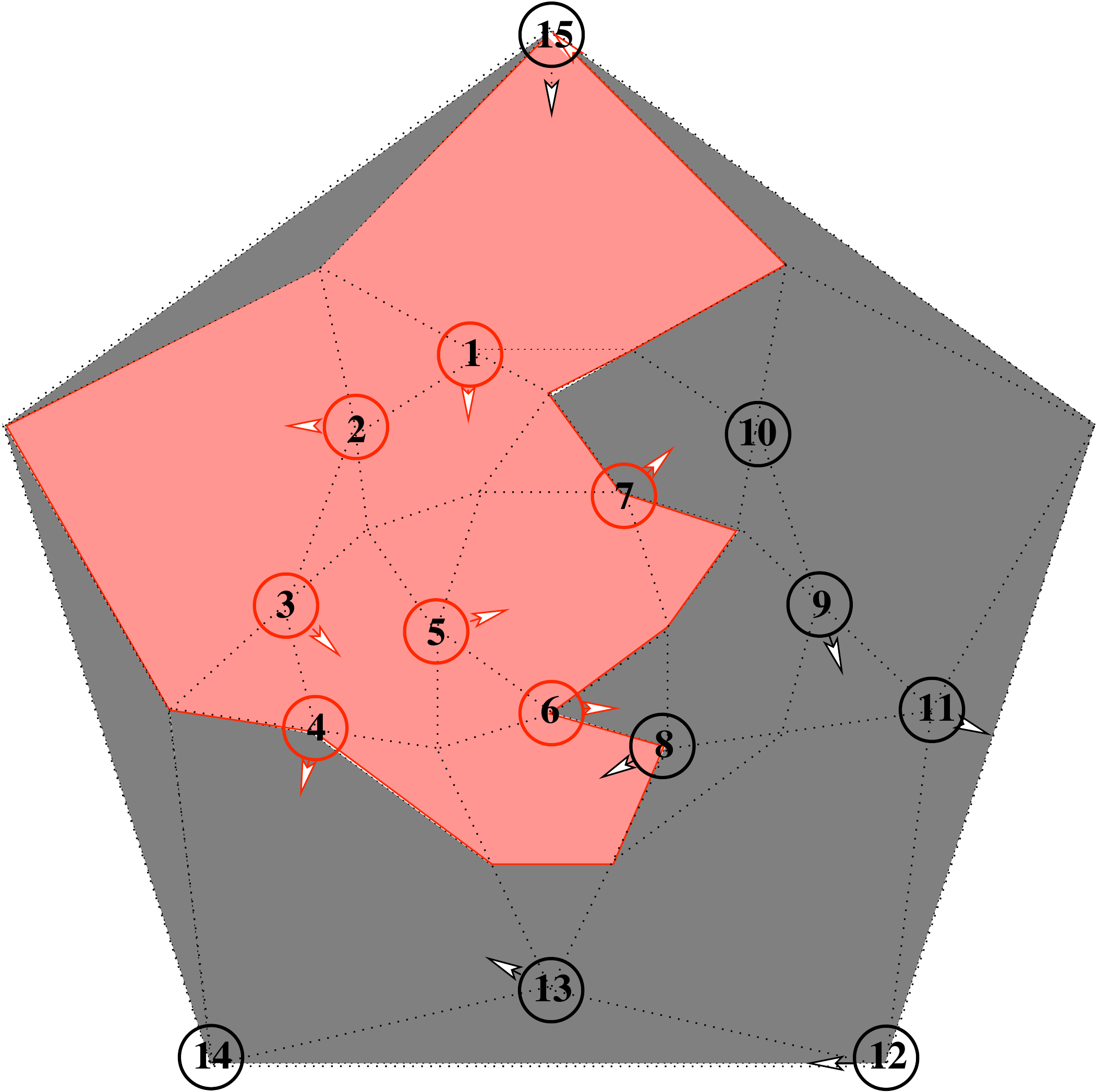}
\end{center}
\caption{{\em A two strand configuration obtained after fifteen vertex replacements.}}
\label{solutionnocross}
\end{figure}

%In this case, 
{ In the case where none of the sixty edges of the icosidodecahedral cage has an additional cross-over, as illustrated for one edge in} %which corresponds to 
Fig.~\ref{options}(a), the start configuration consists of thirty-two loops corresponding to the thirty-two faces of the polyhedron, { and all 4-junctions (located at the thirty vertices of the cage) are of Type A as in Fig.~\ref{replace}(a)} . Since every { 4-junction replacement in Fig.~\ref{replace}(b)} reduces the overall number of loops in the start configuration by two, the minimal number of circular strands needed to realize the cage is two and occurs after fifteen replacements. There is a plethora of different ways of carrying out these replacements in order to arrive at a configuration with only two independent loops { $L_1$ and $L_2$, where $L_i, i=1,2$ result from merging $n_i$ original start configuration loops,  with $n_1+n_2=32.$} % For example, replacements can be chosen such that the resulting two loops consist of different combinations of loops in the start configuration.
 Extremal cases are { (1)  $n_1=31$ and $n_2=1$, i.e. a large loop $L_1$ combining thirty-one individual loops of the start configuration together with a single loop $L_2$ from the start configuration that can not be combined with $L_1$  via the 4-junction replacement of Fig.~\ref{replace}(b), and (2) $n_1=15$ and $n_2=17$, i.e.  two loops of approximately the same size merging fifteen and seventeen loops respectively in the start configuration. An example of the latter situation is presented in Fig.~\ref{solutionnocross}. Junctions are represented by circles, while the numbers indicate the order in which 4-junction replacements Fig.~\ref{replace}(b) are carried out. In the case where the 4-junction before replacement hosts four independent loops, the 4-junction replacement leads to the merging of loops corresponding to three of the four faces meeting at this particular vertex, and the arrow indicates the face (and hence the loop)  that has not been affected by the replacement.  If instead the 4-junction before replacement only hosts  three independent loops, the replacement 4-junction merges these three loops in a single one, and no arrow is present.} After seven 4-junction replacements, the loops corresponding to all red-shaded faces have been combined into a single loop. Replacements 7 to 15 moreover unite the loops on all remaining faces into a separate loop that covers all grey-shaded faces.  The end result is a template for a duplex cage structure that can be realized by two circular ssDNA molecules. 
 
 { We end by  noting that it is not possible, via the replacement in Fig.~\ref{replace}(b), to create two loops $L_1$ and $L_2$ which merge an equal number of original loops in this case, as the topology forces $L_1$ and $L_2$ to combine  odd numbers of loops from the start configuration. }
%because the combined loops by construction are combinations of odd numbers of loops in the start configuration. 
 %In Fig.~\ref{solutionnocross} we present an example where the resulting two loops cover 15, respectively 17, faces. 
%----------------------------------------- Figure --------------------
%--------------------------------------------------------------------
%Numbers on the junctions indicate the order in which replacements are carried out. Each such replacement leads to an identification of the loops corresponding to three of its incident faces, and the arrow indicates the face that has not been affected by the replacement. After 7 replacements, the loops corresponding to all red-shaded faces have been combined into a single loop. Replacements 7 to 15 moreover unite the loops on all remaining faces into a separate loop that covers all grey-shaded faces.  The end result is a template for a duplex cage structure that can be realized by two circular ssDNA molecules. 

\subsection{All edges have an additional cross-over.}

It may be desirable in an experimental set-up to have a cage structure of a size that requires the occurrence of an additional half-turn on each edge such as in Fig.~\ref{options}(b), and we therefore also investigate this scenario.  

In that case, beads are needed according to the bead rule in order to obtain a start configuration. The minimal number of beads required is easily calculated. All faces of the { icosidodecahedron} have an odd number of sides: there are twelve pentagons and twenty triangles. Each face must have an even number of cross-overs for orientability. So in particular, each of the twenty triangles must receive at least one bead, but by placing a bead on each triangle, one actually places at least one bead on each pentagon. The distribution of this minimum number of  twenty beads should be such that pentagons receive an odd number of beads. Let $\alpha$ be the number of pentagons receiving one bead, $\beta$ be the number of pentagons receiving three beads and $\gamma$ be the number of pentagons receiving five beads. Given that there are twelve pentagons in total, we must satisfy the two equations
\ba 
\alpha+3\beta+5\gamma&=&20,\nonumber\\
\alpha+\beta+\gamma&=&12,
\ea
with $\alpha, \beta$ and $\gamma$ positive integers. There are three solutions to the problem, namely
\ba \label{3cases}
&&{\rm Case\, I}\qquad\qquad \,\,\,\,\alpha=8, \, \beta=4,\, \gamma=0\nonumber\\
&&{\rm Case\, II}\qquad \qquad\,\,\alpha=9, \, \beta=2,  \, \gamma=1\nonumber\\
&&{\rm Case\, III}\qquad \qquad\alpha=10,  \, \beta=0,  \, \gamma=2.
\ea
We call the three options in \eqref{3cases} case I, II and III, respectively, and start by considering case I. 
This tells us that the bead rule is fulfilled if there are four pentagons with three beads each, and if every triangle has precisely one bead. We therefore determine all symmetry-inequivalent start configurations with that property. 

Since this is a significant combinatorial task for the polyhedron under consideration, a purely analytical approach as in \cite{JT} is inappropriate here. We therefore adopt a combined analytical/computational approach. 

In the first instance, we use the icosahedral symmetry to reduce the number of options to be considered. In particular, we determine all symmetry-inequivalent  distributions of four pentagonal faces on the icosidodecahedron. Each of these four faces will then have three of its { sides} decorated by one bead, whilst all other pentagons and all triangles will have only one bead on their perimeter.
In order to determine all inequivalent configurations of four pentagons, we consider the equivalent problem of finding all different possibilities of colouring four of the twelve vertices of an icosahedron.  

\begin{figure}[ht]
\begin{center}
(a)\includegraphics[width=4.5cm,keepaspectratio]{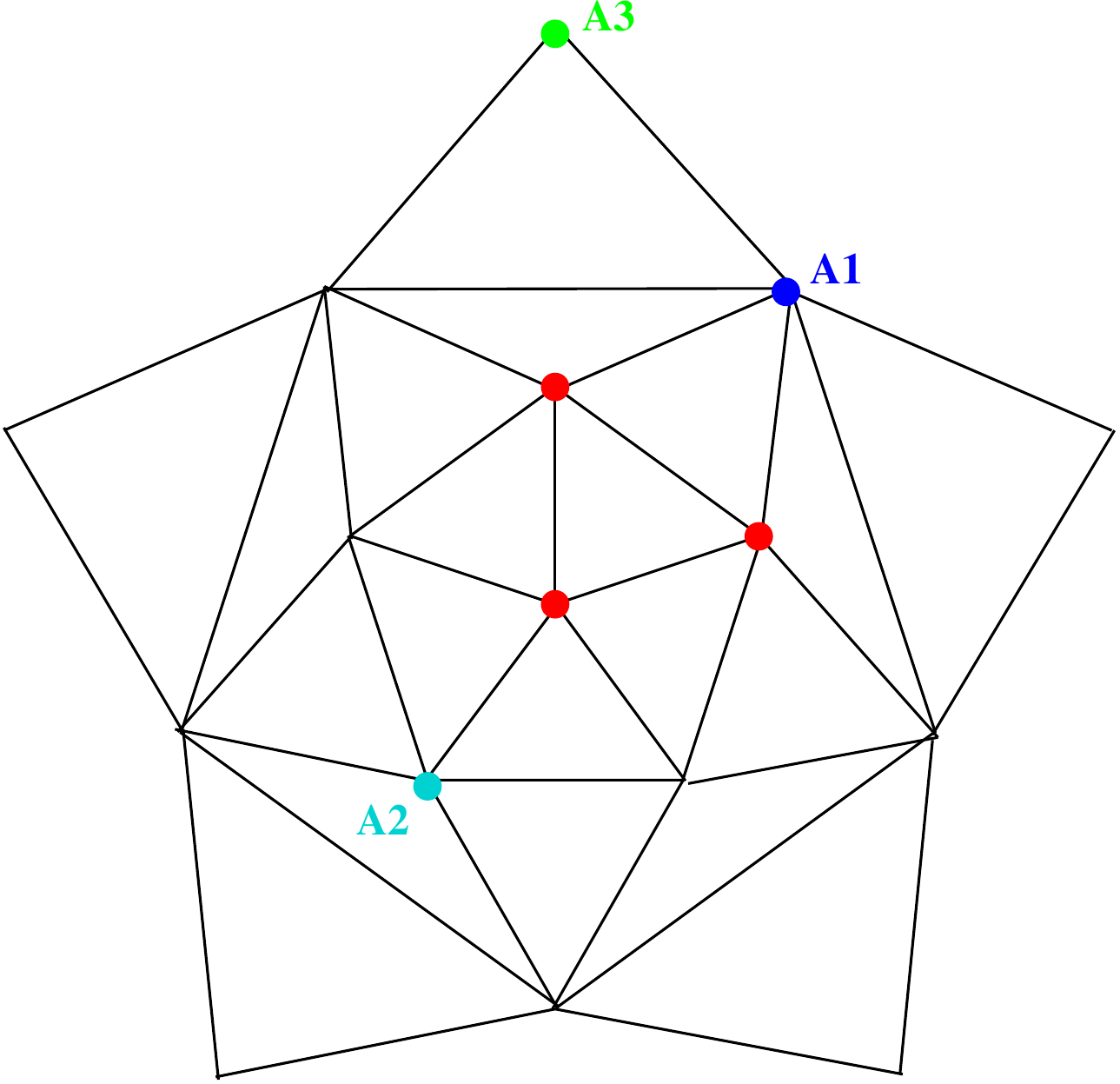}
(b)\includegraphics[width=4.5cm,keepaspectratio]{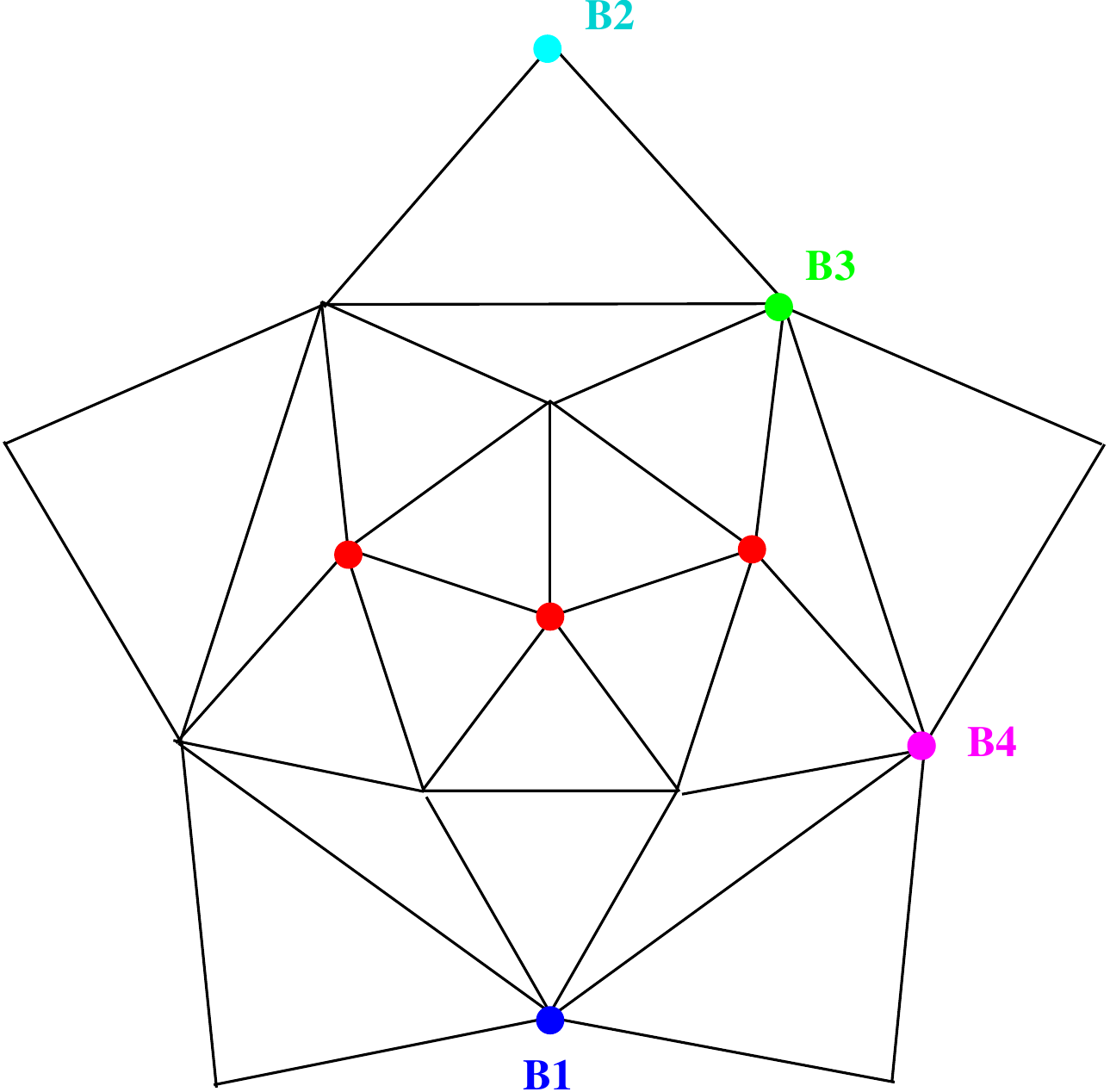}
(c)\includegraphics[width=4.5cm,keepaspectratio]{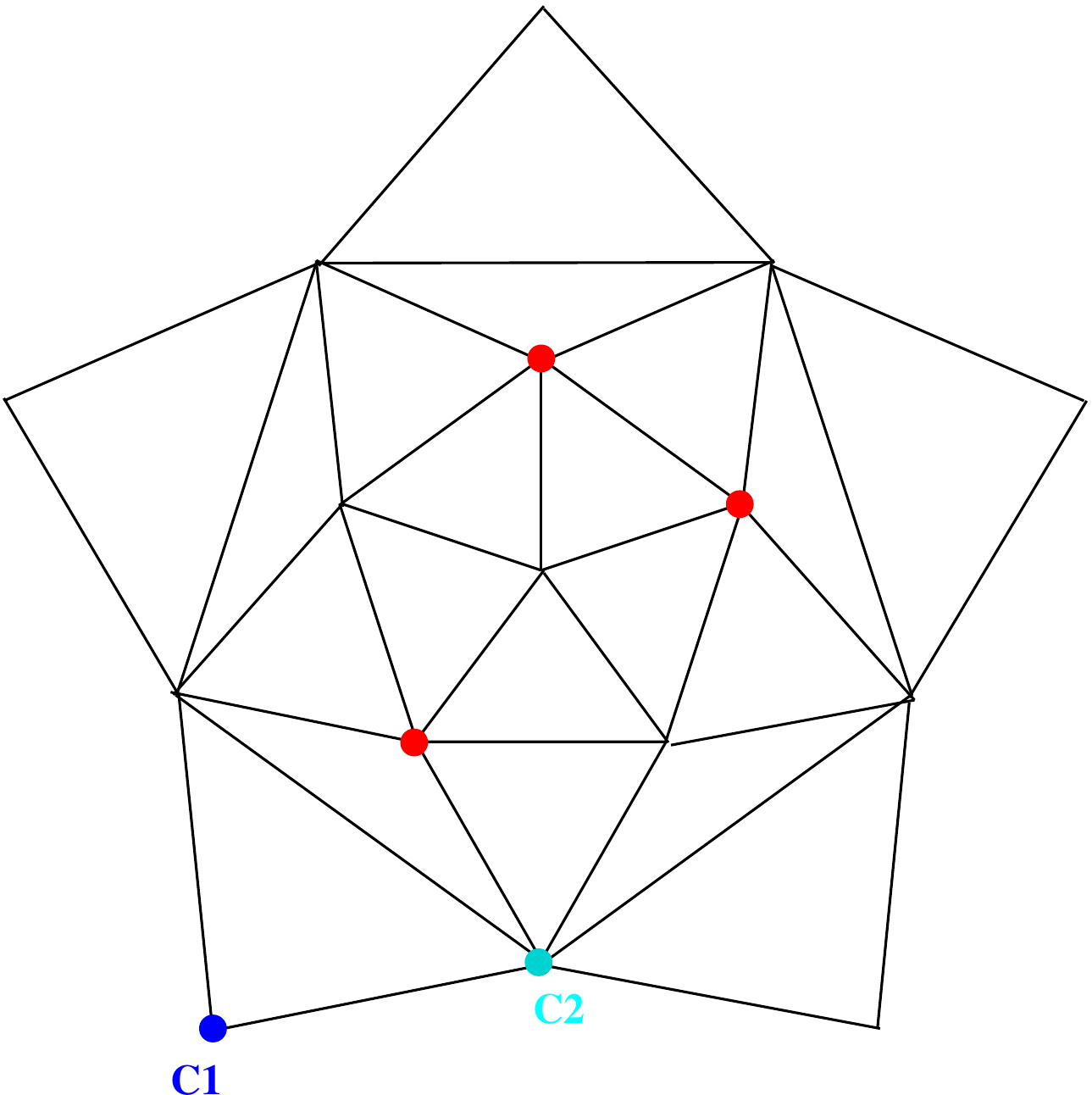}
\end{center}
\caption{{\em Partial start configurations for case I, with a distribution of four out of twelve pentagons on the icosidodecahedron, represented here as distributions of four vertices  on an icosahedron: (a) configurations with three vertices being those of a triangle (red) and the fourth vertex being either $A_1$, $A_2$ or $A_3$; (b) configurations with three red vertices and the fourth one being either $B_1$, $B_2$, $B_3$  or $B_4$; (c) configurations with three red vertices and the fourth one being either $C_1$ or $C_2$.}}
\label{partialstart}
\end{figure}

There are nine inequivalent such configurations for case I, which we call the {\it partial start configurations}. We show them schematically in a projective view of the icosahedron in Fig.~\ref{partialstart}. 

%----------------------------------------- Figure --------------------
%--------------------------------------------------------------------

{\it Inequivalent bead configurations for each partial start configuration}: 
Each partial start configuration encodes several possible cage scenarios which correspond to all inequivalent ways of placing three beads on the { sides} of four distinguished  pentagons, and one bead on one of the sides of all other faces (pentagonal or { triangular}). We carry out this task computationally. 

We have written a computer programme that tests, for each start configuration, which combinations of beads are possible, given the fact that the four distinguished pentagonal faces each have three beads, while all others have one. The results of this programme are encoded as vectors with twenty entries, where each entry represents a triangular face of the icosidodecahedron. The entries are from the set $\{0,1,2\}$, and encode which of the three { sides} contains the bead with respect to our labelling system. In a next step, we translate each such vector into a configuration of loops as follows: we start with an arbitrary edge and, following the rules implied by crosses and beads, continue until we meet the starting point { and therefore form a loop}. We then choose another starting point and follow the same scenario until the entire graph has been covered in loops, such as in Fig.~\ref{initialdata} for example. We then perform a similar analysis for cases II and III. 

The results of this approach are summarised in Table \ref{tab1}. 
\begin{table}[h]
\centering
\begin{tabular}{|c | c | c | c |}
\hline
Loop number & Case I & Case II & Case III \\ \hline
10 & 11527 & 0 & 0\\
12 & 343 & 951 & 0\\
14 & 3 & 8 & 73 \\ 
16 & 0 & 0 & 1 \\ \hline
 \end{tabular}
\caption{\em The distribution of 10-, 12-, 14- and 16-loop start configurations among the three different cases.}
\label{tab1}
\end{table}
In particular, the smallest number of distinct loops is ten and the largest number sixteen. While there are over $10^4$ distinct 10-loop configurations, there is only one configuration with sixteen loops, occurring in case III. The vertex configurations involve either four, three, or two distinct loops.  We use a computer programme to determine the occurrence of the different vertex types for each start configuration. Remarkably, for all start configurations with ten loops, at most three distinct loops meet at each junction. Since {  replacements of type Fig.~\ref{replace}(b) reduce the total number of strands by two at each incidence}, and hence by an even number in total,  the minimal number of loops needed to realise the cage structure via our formalism is two. These two loops could potentially be { merged} %identified 
as in \cite{JT} by a hairpin construction, but this is not considered here. 

\begin{figure}[ht]
\begin{center}
(a)\includegraphics[width=6cm,keepaspectratio]{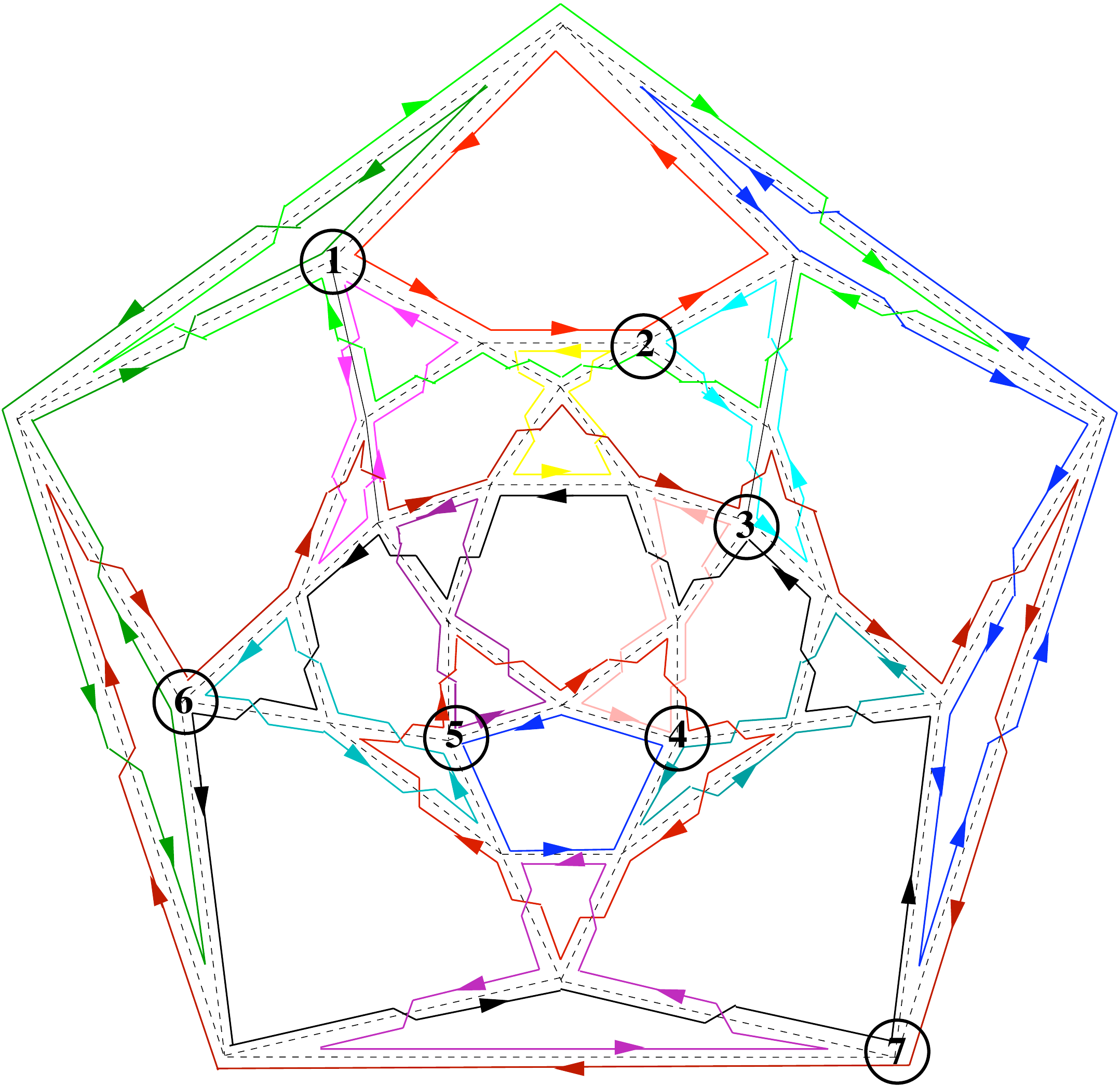}\qquad\qquad
(b)\includegraphics[width=6cm,keepaspectratio]{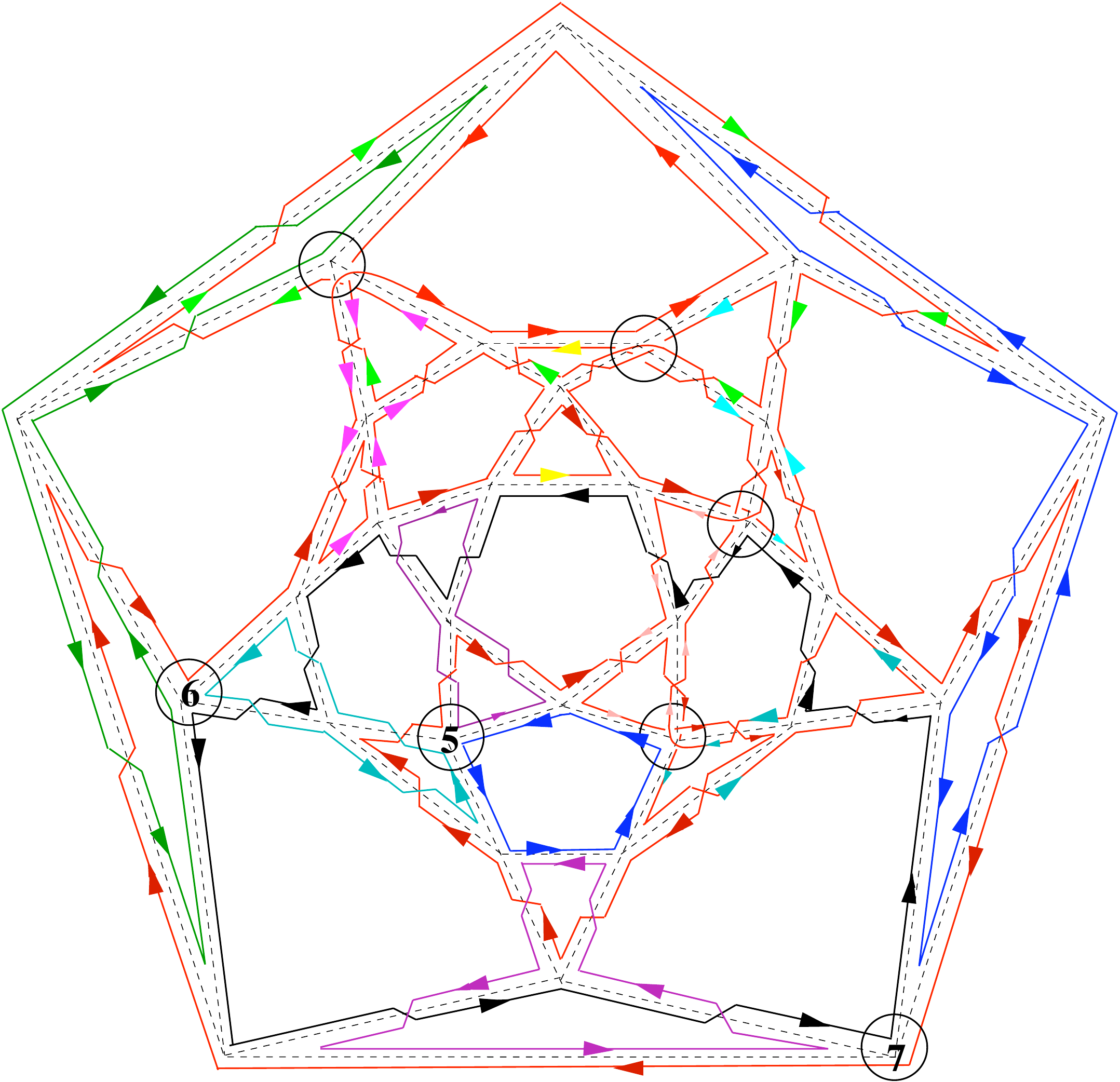}\\
\vskip1cm
(c)\includegraphics[width=6cm,keepaspectratio]{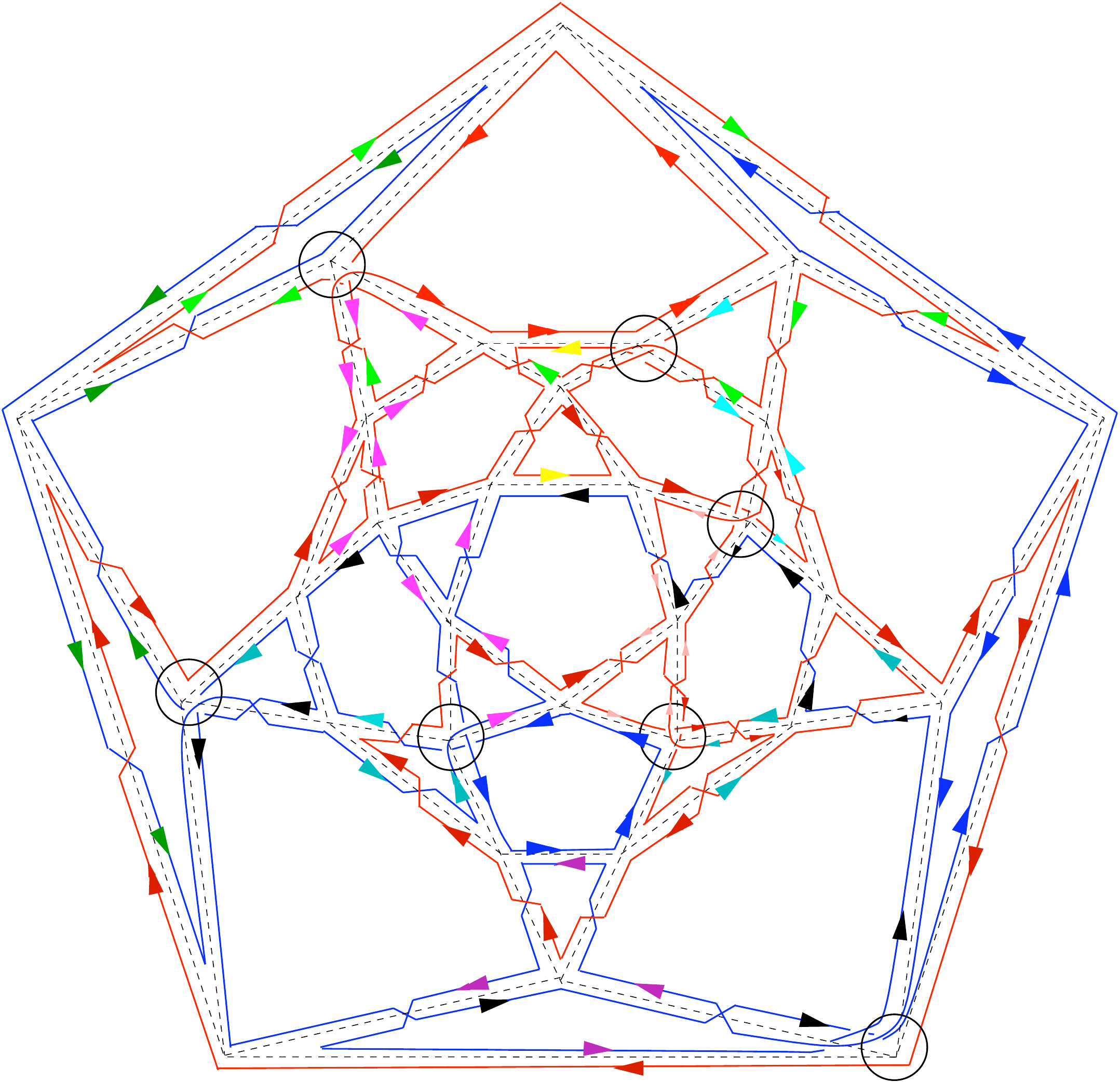}
\end{center}
\caption{{\em The 16-loop start configuration with numbered junctions indicating the replacements that lead to two distinct loops obtained from merging seven (resp. nine) starting configuration loops.}}
\label{16loops}
\end{figure}

Since there is only one start configuration with sixteen loops, we consider this case first: seven replacements are required to reduce the overall number of loops to two, and there are many different ways of achieving this.  In Fig.~\ref{16loops} we display one solution that leads to two loops { $L_1$ and $L_2$ with $n_1=7$ and $n_2=9$.}%uniting 7, respectively 9, of the loops in the start configuration. 
%----------------------------------------- Figure --------------------
%--------------------------------------------------------------------
{ In Fig.~\ref{16loops}(a), circles  indicate the locations of the junctions at which replacements are taking place and the numbers  keep track of the order of replacements by junctions of type Fig.~\ref{replace}(b).} %via numbered circles. 
Fig.~\ref{16loops}(b) shows the result after  replacements 1 to 4 have been carried out to merge nine loops into one larger loop (shown in red). Colours of the arrows on the strands indicate the colour of the loop before it has been merged into the larger (red) loop via the 4-junction replacement. Fig.~\ref{16loops}(c) presents the resulting configuration in terms of two { independent} %separate 
loops, { the red one being that of  Fig.~\ref{16loops}(b)),  and the blue one being obtained after replacements 5 to 7). This} provides a template for the realisation of the cage in terms of two circular ssDNA molecules.

\begin{figure}[ht]
\begin{center}
\includegraphics[width=8cm,keepaspectratio]{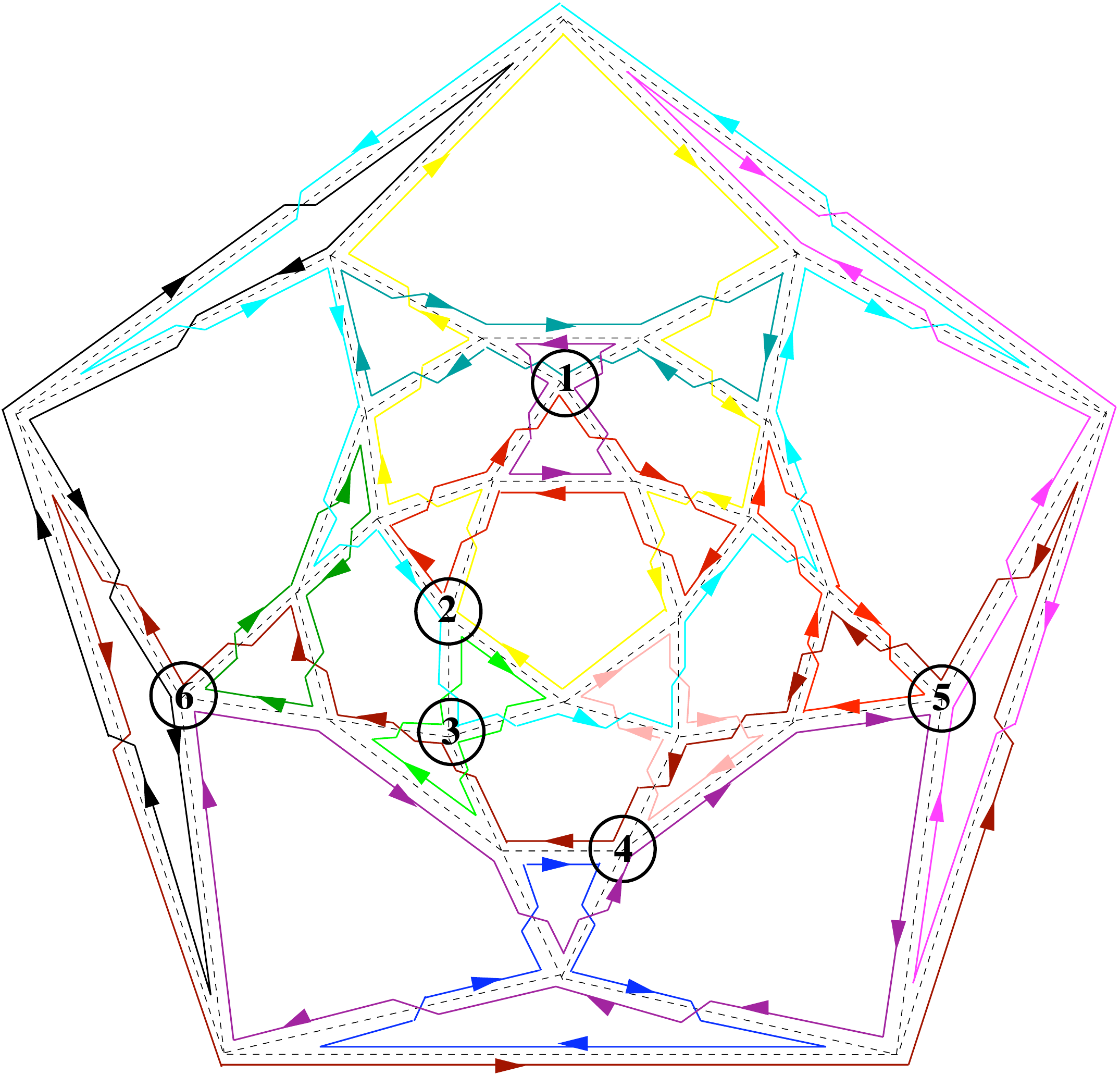}
\end{center}
\caption{{\em A 14-loop starting configuration with numbered junctions indicating the replacements that lead to two distinct loops obtained from merging seven starting configuration loops each.}}
\label{14loops}
\end{figure}

Among the start configurations with twelve or fourteen loops, there are various occurrences of vertices with four distinct loops, for example either none or in-between three and seventeen such vertices for the 12-loop configurations, and either none or in-between ten and eighteen such vertices for the 14-loop configurations. As an example we consider the 14-loop configuration in Fig.~\ref{14loops}.
%----------------------------------------- Figure --------------------
%--------------------------------------------------------------------
In this case, six replacements are necessary to obtain a two-loop configuration. We indicate the locations of six replacements leading to a configuration in terms of two separate loops, each uniting seven loops of the start configuration { (i.e $L_1$ and $L_2$ with $n_1=n_2=7$)}. Note that unlike the previous case { of a 16-loop start configuration}, a 14-loop start configuration can be realised in terms of two loops { merging} %covering 
the same number of smaller loops in the start configuration, because { 14 is twice an odd number, and loops $L_i$ can only merge odd numbers of starting configuration loops. In general, it is only possible to obtain two loops $L_1$ and $L_2$ with $n_1=n_2$ if the start configuration exhibits $2(2k+1)$ loops for $k$ integer (and then, $n_1=n_2=2k+1$). So it is possible for the 10-loop start configurations, but not for any other start configuration corresponding to an entry in Table~\ref{tab1}.}%the number of loops combined is odd. 

The configurations in Fig.~\ref{16loops}(c) and Fig.~\ref{14loops} are two examples of cage structures that can be realized via two circular DNA strands. In contrast with the cage in Fig.~\ref{solutionnocross}, however, there are extra stresses created by the twists that have been introduced in order to make the structure orientable. These stresses can be compensated in two ways, either by introducing a few additional nucleotides in the non-basepaired middle portion of the junction on the expense of making these junctions slightly less rigid, or, if rigid junctions are wanted, by adjusting the lengths of the edges accordingly. A given experimental setting or desired application would dictate which of these options is more appropriate.

\section{Discussion}

We have performed a theoretical analysis of icosidodecahedral cages formed from two circular DNA molecules in a duplex structure. With applications in nanotechnology in mind, emphasis was placed on minimising mechanical stress at the vertex junctions. We have shown that as for the dodecahedral cages considered in \cite{JT}, there exist realisations of the icosidodecahedral cage in terms of two DNA molecules. However, the icosidodecahedral cages considered here have a larger volume per surface ratio than the dodecahedral cages and may therefore be more suitable for nanotechnology applications in which the cages serve as containers for storage or the transport of a cargo. 

Various types of crystallographic cages have been realised before, and we hope that the blueprints for the organisation of icosidodecahedral cages suggested here may assist in their experimental realisation. In particular, these blueprints suggest the structures of the junction molecules that may be used as basic building blocks for the self-assembly of those cages along the lines of \cite{JACS,AJS}.

\section*{Acknowledgements}
AT has been supported by an EPSRC Springboard Fellowship, and RT by an EPSRC Advanced Research Fellowship. RT moreover gratefully acknowledges funding for NG via a Research Leadership Award of the Leverhulme Trust. 

%%%%%%%%%%%%%%%%%%%%%%%%%%%%%%%%%%%%%%%%%%%%%%%%%%
%%%%%%  Bibliography  %%%%%%%%%%%%%%%%%%%%%%%%%%%%
%%%%%%%%%%%%%%%%%%%%%%%%%%%%%%%%%%%%%%%%%%%%%%%%%%

\end{document}